\documentclass[8.5pt,twoside,twocolumn]{article}
\usepackage[super,sort&compress,comma]{natbib} 
\usepackage{mhchem}
\usepackage{times}
\usepackage{amsmath, amssymb, mathptmx} 
\usepackage{sectsty}
\usepackage{balance} 
\usepackage{graphicx,epsfig,array} 
\usepackage{lastpage}
\usepackage[format=plain,singlelinecheck=false,font=small,labelfont=bf,labelsep=space]{caption} 
\usepackage{fancyhdr}
\usepackage{color}

\usepackage{float}
\usepackage{amsmath}
\usepackage{amsfonts}
\usepackage{graphicx}
\usepackage{subcaption}
\usepackage{parskip}
\usepackage{algorithmicx}
\usepackage{algpseudocode}
\usepackage{algorithm}
\usepackage{circuitikz}
\usepackage{ulem}

\newcommand{\mytensor}[1]{\mathbf{#1}}

\newcommand{\tr}[0]{\mathrm{tr}}

\begin{document}


\makeatletter 
\def\subsubsection{\@startsection{subsubsection}{3}{10pt}{-1.25ex plus -1ex minus -.1ex}{0ex plus 0ex}{\normalsize\bf}} 
\def\paragraph{\@startsection{paragraph}{4}{10pt}{-1.25ex plus -1ex minus -.1ex}{0ex plus 0ex}{\normalsize\textit}} 
\renewcommand\@biblabel[1]{#1}            
\renewcommand\@makefntext[1]%
{\noindent\makebox[0pt][r]{\@thefnmark\,}#1}
\makeatother 
\renewcommand{\figurename}{\small{Fig.}~}
\sectionfont{\large}
\subsectionfont{\normalsize} 

\renewcommand{\headrulewidth}{1pt} 
\renewcommand{\footrulewidth}{1pt}
\setlength{\arrayrulewidth}{1pt}
\setlength{\columnsep}{6.5mm}
\setlength\bibsep{1pt}

\twocolumn[
  \begin{@twocolumnfalse}
    \noindent\LARGE{\textbf{Modeling Tensorial Conductivity of Particle Suspension Networks}
    }
    
    \vspace{0.6cm}
    
    \noindent\large{\textbf{Tyler Olsen\textit{$^{a}$} and Ken Kamrin$^{\ast}$\textit{$^{a}$}}}\vspace{0.5cm}
    
    \noindent\textit{\small{\textbf{Received Xth XXXXXXXXXX 20XX, Accepted Xth XXXXXXXXX 20XX\newline
          First published on the web Xth XXXXXXXXXX 20XX}}}
    
    \noindent \textbf{\small{DOI: 10.1039/b000000x}}
    \vspace{0.6cm}
    
    %
    
    \noindent \normalsize{Significant microstructural anisotropy is known to develop during shearing flow of attractive particle suspensions.  
      These suspensions, and their capacity to form conductive networks, play a key role in flow-battery 
      technology, among other applications.
      Herein, we present and test an analytical model for the tensorial conductivity of attractive particle suspensions.   
      The model utilizes the mean fabric of the network to characterize the structure, and the 
      relationship to the conductivity is inspired by a lattice argument.  
      We test the accuracy of our model against a large number of computer-generated suspension networks, 
      based on multiple in-house generation protocols, giving rise to particle networks that emulate the physical system. 
      The model is shown to adequately capture the tensorial conductivity, both in terms of its invariants and its mean directionality.}
    \vspace{0.5cm}
  \end{@twocolumnfalse}
]

\section*{Introduction}

\footnotetext{\textit{$^{a}$~Department of Mechanical Engineering, MIT, Cambridge, MA, USA. }}

The electrical conductivity of heterogeneous materials has been extensively
studied by many different researchers over the years
\cite{Batchelor1974,Cheng1997,Torquato1982,Torquato1985,Jagota1990}.
The literature primarily focuses on heterogeneous materials which are mixtures of
two materials that each have different, isotropic electrical conductivities.
The most well-known result is that of Maxwell, which is based on an effective-medium
approximation for dilute suspensions \cite{Maxwell1881}.
Hashin and Shtrikman approached the problem in a different way.
Rather than attempt to solve for an exact expression for the effective conductivity
of a randomly structured material, they applied a variational method to derive
upper and lower bounds on the effective conductivity \cite{Hashin1962}.
They chose to use a variational approach to derive bounds on the conductivity
because solving the exact problem for an arbitrarily structured heterogeneous
material was analytically intractable.
Torquato \cite{Torquato1982,Torquato1985,TorquatoBook} has studied the effective conductivity problem
in great depth.  He has improved the bounds laid out by Hashin and Shtrikman, has solved for
effective conductivity of a number of different lattice types, and has
expressed the exact tensorial effective conductivity in terms of an infinite
series of N-point probability functions, which can be used to describe the
microstructure of a heterogeneous material. 
The particular case of a suspension consisting of a conductive particle network within an insulating medium 
has been considered theoretically, to our knowledge, in one existing study \cite{Jagota1990}.
The approach they take assumes a spatially homogeneous potential gradient 
field imposed upon the structure, leading to a model for the conductivity
that can be proven to be an upper bound.

\begin{figure}[t]
  \centering
  (a)    \includegraphics[width=.35\textwidth]{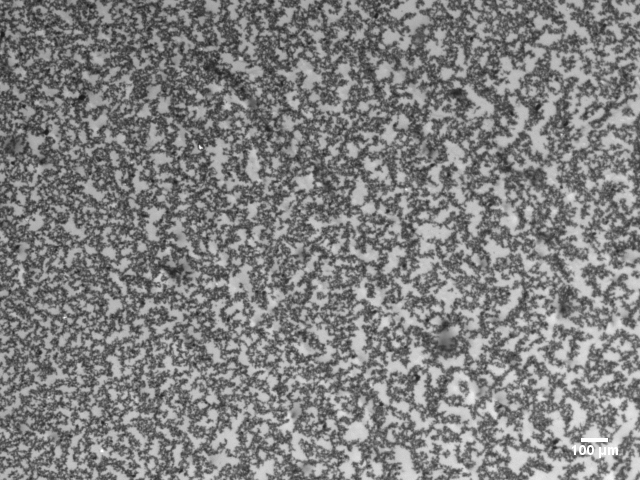}\\
  \vspace{.3cm}
  (b)    \includegraphics[width=.35\textwidth]{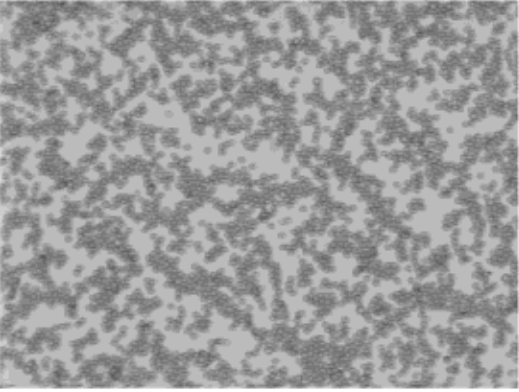}
  \caption{
    (a) Image of a carbon black particle network, an electrically conductive suspension  \cite{Helal2014}. 
    (b) Image of an effective  two-dimenionsional suspension (attractive polystyrene beads on a fluid surface), 
    which has been subjected to shearing. Note the formation of an anisotropic contact network between particles. 
    \cite{Hoekstra2003}
  }  
  \label{fig:AhmedCB} 
  \label{fig:HoekstraPic}
\end{figure}

Much of the aforementioned work is concerned with the isotropic
conductivity of heterogeneous materials. 
In this work, we aim to model the full tensorial conductivity, 
with a focus on suspended networks of conductive particles. 
These particle networks are of practical importance, especially 
in flowable battery technology currently under development by the Joint Center for Energy
Storage Research (JCESR) \cite{Duduta2011}.
In these batteries, a conductive, flowing suspension of carbon black forms
an integral component of the system, see Figure \ref{fig:AhmedCB}(a).
It has been shown in related systems \cite{Hoekstra2003} that shearing flows induce anisotropy
in a contact network of suspended particles, as pictured in Figure \ref{fig:HoekstraPic}(b).
In instances where suspension conductivity arises from particle-particle contacts,
this structure anisotropy should give rise to conductivity anisotropy.
It is this behavior that we seek to describe. 
It has been shown experimentally that the electrical conductivity of a suspension is
highly sensitive to shear rate\cite{Amari1990}, dropping by several orders of magnitude
as shear rate increases.
From this observation and the evidence of particle microstructure changing in
shearing flow, we deduce that a suitably chosen description of the particle network
should be sufficient to predict the electrical conductivity of a suspension.

In the granular media literature, a great deal of attention has been given
to describing the structure of the contact network between particles. 
Perhaps the simplest structural measure for such a network that includes 
anisotropy is the \textit{fabric tensor} \cite{Oda1982,Mehrabadi1982,Satake1978}.  
While more complex structural measures exist, such as pair- and higher-order particle
correlation functions\cite{Torquato1982},  
whose use could enable greater accuracy in constructing a conductivity model, 
we shall show that a suitable model can be achieved solely in terms of the fabric.  
Key to our model development is the solution of a simple case, based on a network 
conforming to a lattice structure.  
The results instruct the form for a new conductivity model, whose accuracy is then 
tested against many thousands of random particle networks.  
To explore a range of particle networks, we describe two distinct algorithms for creating 
random packings --- one for denser packings, and one for more dilute packings that
closely resemble those formed by carbon-black --- and demonstrate the model's predictive
capability against thousands of packings generated from both algorithms.

\section*{Homogenization}
The tensorial form of Ohm's law relates the  electric field vector $\mathbf{E}$ to the current density vector  $\mytensor{J}$ 
through a second-order conductivity
tensor $\mytensor{K}$, i.e.
\begin{equation}\label{eq:OhmVector}
  \mytensor{J} = \mytensor{KE} 
\end{equation}
The conductivity tensor is a symmetric,
positive-definite tensor \cite{TorquatoBook}. An effective conductivity for a representative volume $\Omega$ of a heterogeneous material
must be defined prior to any analytical or numerical work.
The effective conductivity of an ergodic medium is defined by
\begin{equation}
  \label{eq:TorquatoEffective}
 \langle \mytensor{J}\rangle   = \mytensor{K} \langle \mytensor{E}\rangle  
\end{equation}
where $\langle \mytensor{E}\rangle $ and $\langle \mytensor{J}\rangle $ are, respectively, 
the spatially-averaged electric and current density fields over $\Omega$ \cite{TorquatoBook}. 
To avoid a possibly over-reaching assumption of ergodicity --- our tests will be 
conducted on finite domains --- we specify that $\langle \mytensor{E}\rangle $ is imposed by prescribing a linear 
boundary potential ${\varphi(\mathbf{x}\in\partial\Omega)=-\langle\mytensor{E}\rangle\cdot\mathbf{x}}$, 
and that $\langle \mytensor{J}\rangle $ is redefined as the flux that is power-conjugate to $\langle \mytensor{E}\rangle $. 
That is,
\begin{equation}
  \label{eq:PowerDissipation}
 \langle \mytensor{E}\rangle   \cdot  \langle \mytensor{J}\rangle   \equiv  
 \frac{1}{V}\int\limits_\Omega -\nabla\varphi \cdot \mathbf{j} \, \mathrm{dV}
\end{equation}
where $\mathbf{j}$ is the local current density field.  In the ergodic limit of the ensuing analysis, $\langle \mytensor{J}\rangle $ reduces to a standard spatial average. 

Assuming that the current density obeys Kirchoff's current law and Ohm's law ---  
respectively, $\nabla\cdot\mathbf{j}=0$ and $\mathbf{j}=-\sigma \nabla\varphi$ 
for some non-negative conductivity field $\sigma(\mathbf{x})$ --- a symmetric, 
positive-definite conductivity tensor $\mytensor{K}$ must exist that obeys \eqref{eq:TorquatoEffective}.   By using a calculus identity, Eq \ref{eq:PowerDissipation} can be transformed into
\begin{equation}
  \label{eq:PowerDissipation2}
  \langle \mytensor{E}\rangle   \cdot \mytensor{K} \langle \mytensor{E}\rangle   =
  \frac{1}{V}\int\limits_{\partial\Omega} -\varphi\mytensor{j}\cdot\mytensor{n}\,\mathrm{dA}
\end{equation}
where $\mytensor{n}$ is the outward-pointing normal vector. 

We model the particles as perfect conductors, the fluid as a perfect insulator, 
and we suppose electrical resistance arises only at the contacts between particles.
Likewise, the field $\varphi$ is approximated as a constant within each particle but 
possibly varying from particle to particle. 
The above integral can now  be broken into a sum of integrals over the boundary.
In the locations where the boundary passes through free space (i.e., not a particle),
then we know that $\mytensor{j}$ is exactly $\mytensor{0}$.
This leaves only the parts of the boundary that pass through particles, which allows
us to write the integral over the set of boundary particles $B$, i.e.
\begin{equation}
  \label{eq:IntegralSummation}
  \langle \mytensor{E}\rangle  \cdot \mytensor{K}  \langle \mytensor{E}\rangle   =
  \frac{1}{V}\sum\limits_{i\in B}-\varphi_i\int\limits_{\partial\Omega_i} \mytensor{j}\cdot\mytensor{n}\,\mathrm{dA}.
\end{equation}
where $\Omega_i$ is the intersection of the $i$th boundary particle with $\partial \Omega$, 
and the potential within particle $i$, denoted $\varphi_i$ above, can be brought outside 
the integral since it is constant within a particle.
Although the precise nature of $\mytensor{j}$ is unknown within the particle,
the value of the integral $\int_{\partial\Omega_i}\mytensor{j}\cdot\mytensor{n}\,\mathrm{dA}$ 
is the current that is flowing out of $\Omega$.
Denoting this current as $I^{out}_i$ we can write the final expression for the right-hand-side of 
\eqref{eq:PowerDissipation2},
\begin{equation}
  \label{eq:PowerSummation}
  \langle \mytensor{E}\rangle  \cdot \mytensor{K}  \langle \mytensor{E}\rangle  =
  \frac{1}{V}\sum\limits_{i\in B}-\varphi_i I^{out}_i.
\end{equation}
The three independent components of $\mytensor{K}_e$ can be determined by performing
multiple simulations on the same particle network with three non-colinear choices of $\langle \mytensor{E}\rangle $.

\begin{figure}[t]
  \centering
  \includegraphics[width=3in]{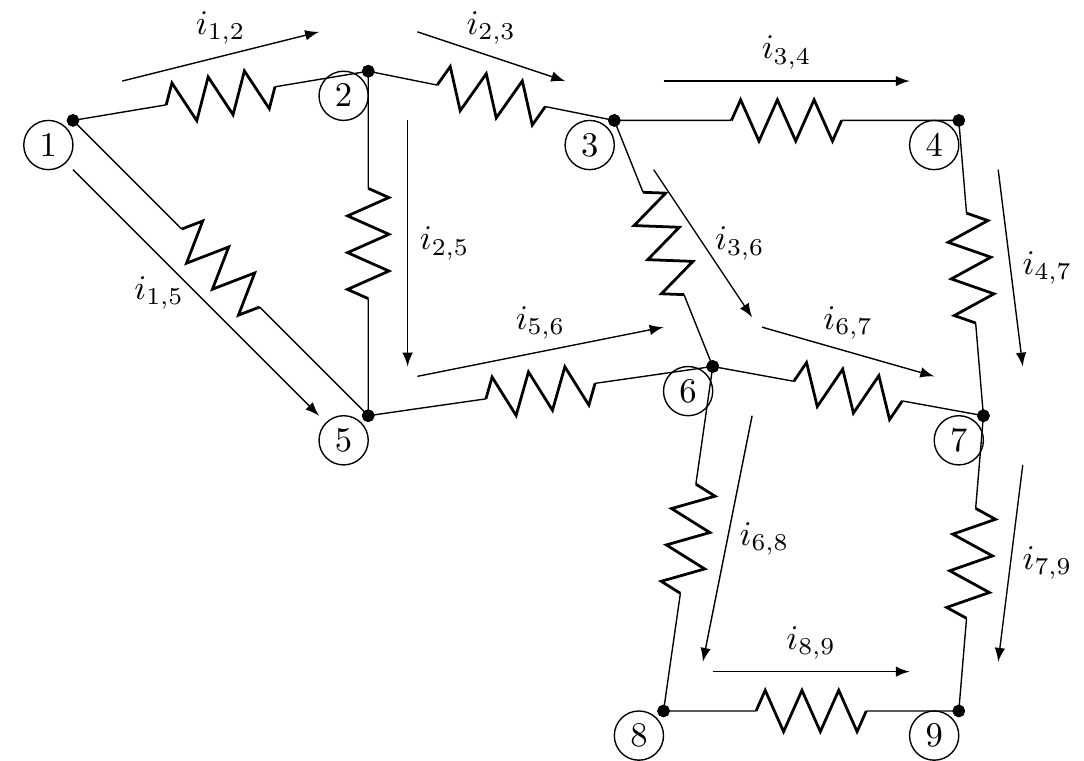}
  \caption{Schematic of a small resistor network with nodes and edges labeled according to our conventions.}
  \label{Schematic}
\end{figure}

By our assumptions for the particle properties, the problem can be reduced 
further to that of a resistor network. 
The network is defined by the set of particles acting as the nodes, which 
are connected by a set of contacts acting as the edges, which carry a resistance $R_c$. 
A schematic of an example network with 9 nodes and 12 edges can be found in figure \ref{Schematic}. 
Supposing an $N$-particle sample and letting $i_{m,n}$ represent the (signed) current 
flowing from particle $m$ to $n$, Ohm's and Kirchoff's law can be rewritten in their simpler discrete form,
\begin{equation} \label{OhmsLaw}
  i_{m,n} = \frac{\varphi_m-\varphi_n}{R_c}
\end{equation}
 and
\begin{equation} \label{KCL}
  \sum_{n} i_{m,n} = 0 \ \ \text{for all} \ \ m.
\end{equation}


\begin{figure}
  \centering
  \makeatletter{}\begin{tikzpicture}
  \coordinate (A) at (0,0);
  \coordinate (B) at (0.866,0.5);
  \coordinate (C) at (-0.5,0.866);
  \coordinate (D) at (0,-1);
  \coordinate (E) at (-1.5, 0.886);
  
    \draw (A) circle (0.5);
  \draw[color=gray] (B) circle (0.5);
  \draw[color=gray] (C) circle (0.5);
  \draw[color=gray] (D) circle (0.5);
  \draw[color=gray] (E) circle (0.5);
  
    \draw[-latex] (A) -> (B) node [near end, above=1pt] {$\mytensor{n}_1$};
  \draw[-latex] (A) -> (C) node [near end, above=2pt] {$\mytensor{n}_2$};
  \draw[-latex] (A) -> (D) node [near end, below=4pt] {$\mytensor{n}_3$};
\end{tikzpicture}
 
  \caption{Schematic of particles in contact showing contact
    vectors $\mytensor{n}_i$.}
  \label{fig:Fabric}
\end{figure}
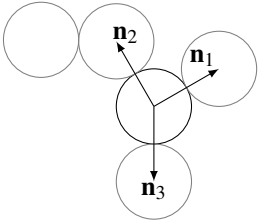
Solving these linear equations for a given particle network enables us to calculate 
$I^{out}$ in (\ref{eq:PowerSummation}) and hence the conductivity tensor for the network.  


We choose to use the fabric tensor as the measure of the network structure.  
The particle-level fabric is a local quantity that can be defined for particle 
$p$ by the relation \cite{Satake1978,Oda1982,Mehrabadi1982} 
\begin{equation}\label{eq:FabricParticle}
  \mytensor{A^p} = \sum_i \mytensor{n}_i \otimes \mytensor{n}_i
\end{equation}
where $\otimes$ denotes the dyadic product, and 
$\mytensor{n}_i$ is the unit normal vector connecting particle centroids of the $i$'th contact on the particle.
This is illustrated in Figure \ref{fig:Fabric}. 
To homogenize over the entire particle network, or at least meso-sized region of it, 
the average fabric tensor is defined as the system average of the particle fabric tensors.
\begin{equation}\label{eq:FabricAverage}
 \mytensor{A} = \frac{1}{N_{\mathrm{particles}}}\sum_{p=1}^{N_{\mathrm{particles}}} \mytensor{A^p}
\end{equation}

The definition of the fabric tensor has some attractive features.  
It is symmetric and positive-semidefinite, guaranteeing that 
the eigenvalues are non-negative and that the eigenvectors are orthogonal.
These properties are shared by the conductivity tensor
$\mytensor{K}$, suggesting the fabric tensor could be an 
appropriate independent variable in the conductivity's functional form.

\section*{Lattice-Reduced Model} 
We propose an analytical model to elucidate the connection between electrical conductivity 
and the fabric tensor based on a simplified lattice structure.  
We will test this model's applicability to random packings in the later sections.  

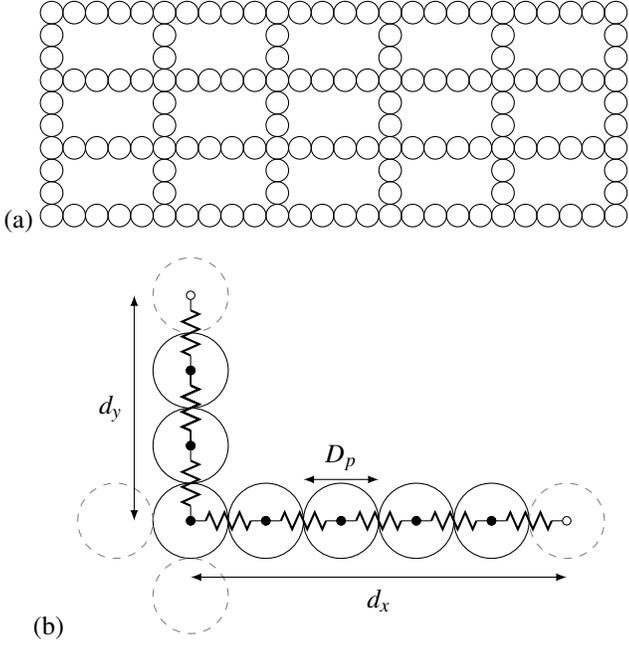
\begin{figure}[t]
  \centering
    \makeatletter{}
  (a)  \begin{tikzpicture}[scale=0.3]
      \draw (0,0) circle (0.5);
      \draw (0,1) circle (0.5);
      \draw (0,2) circle (0.5);
      \draw (0,3) circle (0.5);
      \draw (0,4) circle (0.5);
      \draw (0,5) circle (0.5);
      \draw (0,6) circle (0.5);
      \draw (0,7) circle (0.5);
      \draw (0,8) circle (0.5);
      \draw (0,9) circle (0.5);
      \draw (1,0) circle (0.5);
      \draw (1,3) circle (0.5);
      \draw (1,6) circle (0.5);
      \draw (1,9) circle (0.5);
      \draw (2,0) circle (0.5);
      \draw (2,3) circle (0.5);
      \draw (2,6) circle (0.5);
      \draw (2,9) circle (0.5);
      \draw (3,0) circle (0.5);
      \draw (3,3) circle (0.5);
      \draw (3,6) circle (0.5);
      \draw (3,9) circle (0.5);
      \draw (4,0) circle (0.5);
      \draw (4,3) circle (0.5);
      \draw (4,6) circle (0.5);
      \draw (4,9) circle (0.5);
      \draw (5,0) circle (0.5);
      \draw (5,1) circle (0.5);
      \draw (5,2) circle (0.5);
      \draw (5,3) circle (0.5);
      \draw (5,4) circle (0.5);
      \draw (5,5) circle (0.5);
      \draw (5,6) circle (0.5);
      \draw (5,7) circle (0.5);
      \draw (5,8) circle (0.5);
      \draw (5,9) circle (0.5);
      \draw (6,0) circle (0.5);
      \draw (6,3) circle (0.5);
      \draw (6,6) circle (0.5);
      \draw (6,9) circle (0.5);
      \draw (7,0) circle (0.5);
      \draw (7,3) circle (0.5);
      \draw (7,6) circle (0.5);
      \draw (7,9) circle (0.5);
      \draw (8,0) circle (0.5);
      \draw (8,3) circle (0.5);
      \draw (8,6) circle (0.5);
      \draw (8,9) circle (0.5);
      \draw (9,0) circle (0.5);
      \draw (9,3) circle (0.5);
      \draw (9,6) circle (0.5);
      \draw (9,9) circle (0.5);
      \draw (10,0) circle (0.5);
      \draw (10,1) circle (0.5);
      \draw (10,2) circle (0.5);
      \draw (10,3) circle (0.5);
      \draw (10,4) circle (0.5);
      \draw (10,5) circle (0.5);
      \draw (10,6) circle (0.5);
      \draw (10,7) circle (0.5);
      \draw (10,8) circle (0.5);
      \draw (10,9) circle (0.5);
      \draw (11,0) circle (0.5);
      \draw (11,3) circle (0.5);
      \draw (11,6) circle (0.5);
      \draw (11,9) circle (0.5);
      \draw (12,0) circle (0.5);
      \draw (12,3) circle (0.5);
      \draw (12,6) circle (0.5);
      \draw (12,9) circle (0.5);
      \draw (13,0) circle (0.5);
      \draw (13,3) circle (0.5);
      \draw (13,6) circle (0.5);
      \draw (13,9) circle (0.5);
      \draw (14,0) circle (0.5);
      \draw (14,3) circle (0.5);
      \draw (14,6) circle (0.5);
      \draw (14,9) circle (0.5);
      \draw (15,0) circle (0.5);
      \draw (15,1) circle (0.5);
      \draw (15,2) circle (0.5);
      \draw (15,3) circle (0.5);
      \draw (15,4) circle (0.5);
      \draw (15,5) circle (0.5);
      \draw (15,6) circle (0.5);
      \draw (15,7) circle (0.5);
      \draw (15,8) circle (0.5);
      \draw (15,9) circle (0.5);
      \draw (16,0) circle (0.5);
      \draw (16,3) circle (0.5);
      \draw (16,6) circle (0.5);
      \draw (16,9) circle (0.5);
      \draw (17,0) circle (0.5);
      \draw (17,3) circle (0.5);
      \draw (17,6) circle (0.5);
      \draw (17,9) circle (0.5);
      \draw (18,0) circle (0.5);
      \draw (18,3) circle (0.5);
      \draw (18,6) circle (0.5);
      \draw (18,9) circle (0.5);
      \draw (19,0) circle (0.5);
      \draw (19,3) circle (0.5);
      \draw (19,6) circle (0.5);
      \draw (19,9) circle (0.5);
      \draw (20,0) circle (0.5);
      \draw (20,1) circle (0.5);
      \draw (20,2) circle (0.5);
      \draw (20,3) circle (0.5);
      \draw (20,4) circle (0.5);
      \draw (20,5) circle (0.5);
      \draw (20,6) circle (0.5);
      \draw (20,7) circle (0.5);
      \draw (20,8) circle (0.5);
      \draw (20,9) circle (0.5);
      \draw (21,0) circle (0.5);
      \draw (21,3) circle (0.5);
      \draw (21,6) circle (0.5);
      \draw (21,9) circle (0.5);
      \draw (22,0) circle (0.5);
      \draw (22,3) circle (0.5);
      \draw (22,6) circle (0.5);
      \draw (22,9) circle (0.5);
      \draw (23,0) circle (0.5);
      \draw (23,3) circle (0.5);
      \draw (23,6) circle (0.5);
      \draw (23,9) circle (0.5);
      \draw (24,0) circle (0.5);
      \draw (24,3) circle (0.5);
      \draw (24,6) circle (0.5);
      \draw (24,9) circle (0.5);
      \draw (25,0) circle (0.5);
      \draw (25,1) circle (0.5);
      \draw (25,2) circle (0.5);
      \draw (25,3) circle (0.5);
      \draw (25,4) circle (0.5);
      \draw (25,5) circle (0.5);
      \draw (25,6) circle (0.5);
      \draw (25,7) circle (0.5);
      \draw (25,8) circle (0.5);
      \draw (25,9) circle (0.5);
    \end{tikzpicture} 
 \vspace{.3cm}
    \centering
    \ctikzset{bipoles/length=.75cm}
(b)    \begin{circuitikz}
      \coordinate (A) at (0,0);
      \coordinate (B) at (1,0);
      \coordinate (C) at (2,0);
      \coordinate (D) at (3,0);
      \coordinate (E) at (4,0);
      \coordinate (F) at (0,1);
      \coordinate (G) at (0,2);
      \coordinate (DX) at (0,0);
      \draw (A) circle (0.5);
      \draw (B) circle (0.5);
      \draw (C) circle (0.5);
      \draw (D) circle (0.5);
      \draw (E) circle (0.5);
      \draw (F) circle (0.5);
      \draw (G) circle (0.5);
      \draw[dashed,gray] ($ (G)+(0,1) $) circle (0.5);
      \draw[dashed,gray] ($ (E)+(1,0) $) circle (0.5);
      \draw[dashed,gray] ($ (A)-(1,0) $) circle (0.5);
      \draw[dashed,gray] ($ (A)-(0,1) $) circle (0.5);
      \draw[latex-latex] (0,-0.75) -> (5,-0.75) node [pos=0.5,below=1pt] {$d_x$};
      \draw[latex-latex] (-0.75,0) -> (-0.75,3) node [pos=0.5,left=1pt] {$d_y$};
      \draw[latex-latex] (1.5,0.55) -> (2.5,0.55) node [pos=0.5,above=0pt] {$D_p$};
      \draw
      ($ (A)+(DX) $) to[R,*-*] ($ (B)+(DX) $)
      ($ (B)+(DX) $) to[R,*-*] ($ (C)+(DX) $)
      ($ (C)+(DX) $) to[R,*-*] ($ (D)+(DX) $)
      ($ (D)+(DX) $) to[R,*-*] ($ (E)+(DX) $)
      ($ (A)+(DX) $) to[R,*-*] ($ (F)+(DX) $)
      ($ (F)+(DX) $) to[R,*-*] ($ (G)+(DX) $)
      ($ (F)+(DX) $) to[R,*-*] ($ (G)+(DX) $)
      ($ (E)+(DX) $) to[R,*-o] ($ (E)+(DX)+(1,0) $)
      ($ (G)+(DX) $) to[R,*-o] ($ (G)+(DX)+(0,1) $)
      ;
    \end{circuitikz}
    %
    \caption{
      Idealized particle lattice and unit cell from which fabric-conductivity relation was derived. 
      (a) Example 2D idealized particle lattice. 
      (b) 2D lattice unit cell and its resistor network analog. 
      Neighboring unit cells are shown in gray dashed lines.
    } 
    \label{fig:2DLattice} 
    \label{fig:UnitCell}
\end{figure}

The particles are imagined to live on an idealized infinite, periodic lattice. 
The lattice is parameterized by a set of numbers that describe the particle size and spacing. 
These parameters are 
(1) particle diameter $D_p$,
(2) distance in x-direction between chains $d_x$,
(3) distance in y-direction between chains $d_y$,
(4) distance in z-direction between chains $d_z$.
In 2D, only the first three parameters are used.
An illustration of a 2D lattice characterized by these parameters
is shown in figure \ref{fig:2DLattice}(a), with its fundamental unit cell shown in
figure \ref{fig:UnitCell}(b).
 
%
Both the average fabric tensor and effective conductivity can be
computed analytically.
The average fabric tensor is defined as the spatial average of the
fabric tensor for all of the particles in the unit cell and ultimately results in the formua
\begin{equation}\label{eq:FabricSimplified}
  \mytensor{A}  =
  \frac{2}{N_x + N_y - 1}
  \left[
    \begin{array}{cc}
      N_x & 0 \\
      0 & N_y
    \end{array}
    \right]
\end{equation}
%
%
In this expression, the key quantities to recognize are the
number of particles in the x-oriented chain, $N_x=d_x/D_p$, and
the number of particles in the y-oriented chain, $N_y=d_y/D_p$.

Next, the effective conductivity was derived for the unit cell.
To do this, imagine applying an arbitrary voltage difference
across the x-oriented and y-oriented chains separately.  These voltages are $\Delta \varphi_x$ and $\Delta \varphi_y$, respectively. 
effective resistance.
By applying Ohm's law through the corresponding chains, we can
recover the components of the vector form of Ohm's Law shown
in \eqref{eq:OhmVector}.  For example, for the x-oriented chain
\begin{align}
  j_x &= \left( \frac{1}{N_y R_c} \right) (\nabla \varphi)_x 
\end{align}
with $(\nabla \varphi)_x=\Delta \varphi_x/d_x$. 
Due to the geometry of the problem, we know that the off-diagonal
components of the conductivity tensor $\mytensor{K}$ are exactly zero.
Therefore, we can say
\begin{equation}\label{eq:Kxx}
  K_{11} = \frac{1}{N_y R_c}.
\end{equation}
Similarly analysis yields
\begin{equation}\label{eq:Kyy}
  K_{22} = \frac{1}{N_x R_c}.
\end{equation}
Finally, the parameters $N_x$ and $N_y$
can be algebraically eliminated to give the components of $\mytensor{K}$ in terms
of the components of $\mytensor{A}$, yielding the tensorial relationship
%
%
\begin{equation}\label{eq:KAtensor2D}
  \mytensor{K} = \frac{1}{R_c} \frac{\tr\mytensor{A}-2}{\det\mytensor{A}} \mytensor{A}.
\end{equation}
We refer to the formula in \eqref{eq:KAtensor2D} as the ``lattice model''.
A similar analysis can be carried out for a three-dimensional unit cell,
which will yield the following expression for the conductivity tensor,
\begin{equation}\label{eq:KAtensor3D}
  \mytensor{K} = \frac{1}{4D_pR_c}  \frac{\left(\tr\mytensor{A}-2\right)^2}
         {\det\mytensor{A}}\mytensor{A}.
\end{equation}
The formulae above apply when $\tr \mytensor{A}-2$ is non-negative.  
Otherwise the solution is $\mytensor{K}=\mytensor{0}$. 

Despite its inspiration from the lattice structure, there are several 
reasons to consider the applicability of the lattice model to more general particle networks.  
For one, the formula purports codirectionality of the fabric and conductivity, i.e. 
the deviators of the two tensors are aligned, implying that the direction of anisotropy 
of one tensor gives the anisotropy direction of the other, which to a first approximation 
ought to match the behavior of general particle networks.  Second, the results imply that 
conductivity should vanish when $\tr \mytensor{A}<2$, which is sensible more generally 
(though not strictly) because particles in a percolating chain, as needed to conduct 
current across the sample, must have coordination number at least two.  
Above this threshold, conductivity increases with $\tr\mathbf{A}$ in line with one's basic 
intuition for more highly coordinated networks.  

We are aware of one other fabric-based analytical model for conductive particle networks, 
which was developed by Jagota and Hui \cite{Jagota1990}.  
In their work, a uniformity hypothesis is made with regard to the potential gradient, which results in a conductivity model that is fully linear in the fabric tensor,
\begin{equation}
  \label{eq:JH}
  \mytensor{K}=\frac{N_V D_p^2}{4 R_c}\mytensor{A}.
\end{equation}
The above, which can be proven to be an upper-bound on the real conductivity, 
is for a two-dimensional system and $N_V$ is the particle number fraction (per area in 2D).  
This model differs from ours most notably in that the conductivity is not thresholded 
by the coordination number, the formula depends explicitly on the particle area 
fraction as well fabric, and it does not depend on the fabric determinant.

\section*{Numerical Simulation}

In order to perform numerical experiments and determine the generality of the 
lattice model, a large number of random 
particle networks (packings) must be created. 
There are a number of methods to do this already in the granular and particulate 
matter literature. See the references for a broad summary of the currently
available granular packing algorithms\cite{Bagi2005}.
Attractive suspensions have been modeled with the Diffusion-Limited Aggregation (DLA)
model of Witten and Sander\cite{Sander1981}. A common feature of many of the granular statics methods  is that 
they solve force equilibrium equations for a system of particles 
This was not a feature that was required for this study,
so these types of methods were not used, in the interest of saving computational time.
Instead, we developed two methods for creating two-dimensional random
contact networks of particles, and we tested our model against numerous packings 
generated by each method. Both methods allow us to influence the resulting anisotropic structure of the packings.

\noindent{\textit{Algorithm 1:} Our first packing algorithm was designed to create a dense random
contact networks of particles.
This is in contrast to a later algorithm, to be described below,
which created packings that resulted in much lower-density packings.
The dense packings were created by perturbing a 2D hexagonal close-packing
of particles.
This was achieved by placing points into a triangular lattice,
adding random noise to the position of each point, and finally
growing each particle as large as possible such that no particles overlapped.
Anisotropy can be influenced by shearing the points with an affine
transformation $\mytensor{x}' = \mytensor{F}\,\mytensor{x}$ before
growing the radii.
This process is described in pseudocode below (Algorithm \ref{alg:Algo1}).
An example of the resulting packing overlaid by its
analogous resistor network is shown in figure \ref{fig:Algo1_example}.

\begin{algorithm}[H]
  \caption{}
  \begin{algorithmic}
    \State Seed $L\times L$ box with close-packed points
    \State Perturb points with random noise
    \State Move each point to new location $\mytensor{x}'$ by
    $\mytensor{x}' = \mytensor{F}\,\mytensor{x}$
    \While{Not all radii frozen}
    \State Find smallest distance that any particle can grow
    \State Grow all particles by this amount
    \State Freeze radii of particles that come into contact
    \EndWhile
  \end{algorithmic}
  \label{alg:Algo1}
\end{algorithm}

\begin{figure}
  \centering
  \includegraphics{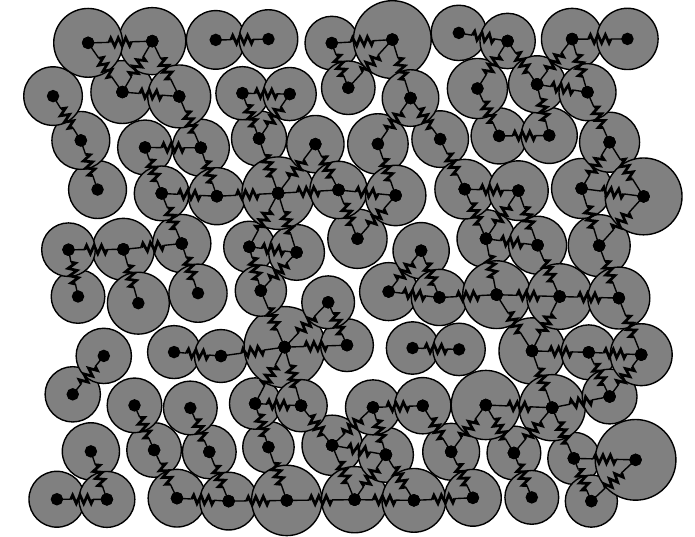}
    \caption{Example (using a small number of particles) of a dense particle packing resulting from Algorithm \ref{alg:Algo1}.}
  \label{fig:Algo1_example}
\end{figure}


\noindent\textit{Algorithm 2:} 
This procedure was motivated by a need to better understand the conductivity of 
carbon black suspensions in an insulating medium. The self-attraction carbon black particles leads to fractal particle networks that are electrically
percolating at low volume fraction (below 1 vol\%)\cite{Duduta2011}.

To produce structures that more closely resemble carbon black suspensions, 
we developed our second packing algorithm, which is inspired by the ``hit-and-stick'' 
behavior of the carbon particles. 
In addition, the new algorithm is able to include the effects of particle Brownian motion but this is not essential to the algorithm.

First, clusters (single particles at this stage) are seeded randomly into a
$L^d$ box, where $d$ is the number of spatial dimensions.
Next, a linear velocity field is imposed directly on each cluster's centroid according to
\begin{equation}\label{eq:velocity}
  \mytensor{v}  = -\mathbf{B}(\mathbf{x}-\mathbf{O})
\end{equation}
where $\mytensor{O}$ is a point in the middle of the original box. 
This imposed velocity field serves to pull all of the clusters together. 
The matrix $\mytensor{B}$ is a $d \times d$ matrix that allows us to impose an anisotropic velocity field. 
This allows us to influence (but not completely impose) the fabric 
tensor that results from this packing method. 
After the velocity field is imposed, the particle positions are updated by assuming a time step dt (computed at runtime). 
Then, the clusters are checked to determine whether any contacts have been made with other clusters. 
If so, the clusters are cohered into a single cluster for all future steps. 
This process of imposing velocity, updating positions, and handling contacts is repeated until only a single cluster remains.  
The process is outlined in pseudocode in Algorithm \ref{alg:Algo2}. 
An example of a packing resulting from this process is shown in figure 
\ref{fig:Algo2_example} and a larger example is displayed in figure \ref{fig:Histogram}. 

The box-counting fractal dimension \cite{falconer2013} of the resulting packings was 
computed in order to determine if they resembled real-life packings found in experiments. 
The fractal dimension of packings produced by this method is approximately $d=1.75$. 
This was compared against the particle network image in figure \ref{fig:HoekstraPic}. 
This network has a fractal dimension of approximately $d=1.7 \pm 0.1$. 
Uncertainty in the measurement is due to the image processing techniques used to identify particles. 
Based on these measurements, we are satisfied that this algorithm produces  
realistic packings, although more detailed correlation function measurements would be needed for a firmer conclusion.

\begin{algorithm}[t]
  \caption{}
  \begin{algorithmic}
    \State Seed N clusters (particles) in $L^d$ square
    \While{$N_{Clusters} > 1$}
    \State Move clusters according to
    $\mathbf{v} = -\mathbf{B}(\mathbf{x}-\mathbf{O})$
    \State Locate collisions between clusters
    \State Combine clusters in contact and recompute centroids
    \EndWhile
  \end{algorithmic}
  \label{alg:Algo2}
\end{algorithm}
\begin{figure}
  \centering
  \includegraphics{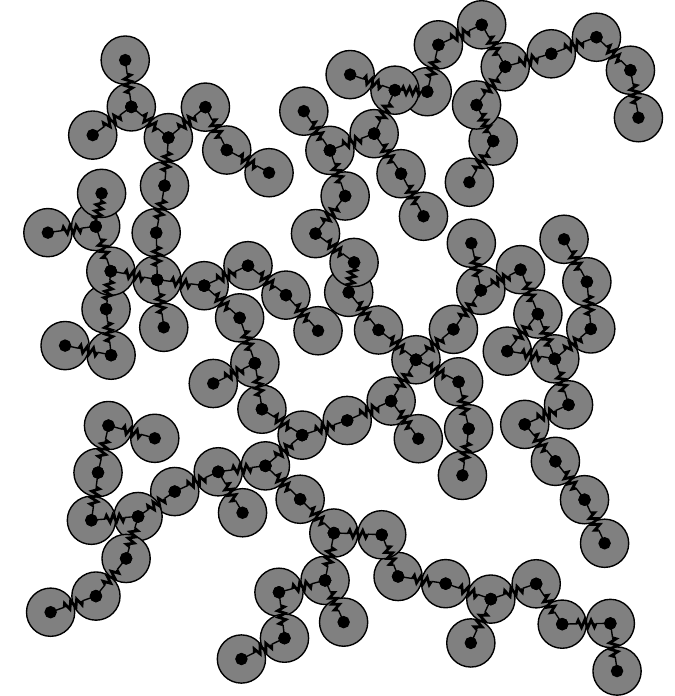}
  \caption{
     Example (using a small number of particles) of a packing resulting from Algorithm \ref{alg:Algo2} using a small number of particles. 
    }
  \label{fig:Algo2_example}
\end{figure}

\noindent\textit{Applying boundary conditions:}  
In order to apply the solution method described above to an arbitrary packing
of particles, appropriate boundary conditions must be applied.
In these simulations, a prescribed voltage was applied to particles all 
around the boundary.
This process consists of two steps: first, the boundary must be identified,
and second, the linear system must be updated to reflect the known voltages.

For the first packing algorithm, identifying the boundary is a trivial process,
since the particle locations are known \emph{a priori}.
For algorithm \ref{alg:Algo2}, however, the particle positions are not known.
A boundary can be located visually quite easily at the end of the simulation
process, but performing this step manually would be prohibitively slow.
In order to expedite and automate the simulation process, the following method
was devised to locate the boundary.

First, histograms of the particle $x$ and $y$ positions were separately created.
To find the ``left'' and ``right'' boundaries, denoted $x^-$ and $x^+$
respectively, the histogram of $x$ positions was thresholded.
The value $x^-$ is defined as the smallest $x$ value where the histogram reaches
$50\%$ of its maximum value.
The value $x^+$ is defined as the largest $x$ value that meets the same criterion.
The top and bottom boundaries, $y^+$ and $y^-$, are found in the same manner
using the histogram of particle $y$ coordinates.
The threshold value $50\%$ was determined emperically to locate the same boundary
that one would identify visually. 
An example packing and its associated $x$-position histogram is shown below in
figure \ref{fig:Histogram} to demonstrate the efficacy of the method.
Once the location of the boundary has been identified, all particles whose
centers fall less than one radius away from the lines are marked as being
``boundary particles''.

\begin{figure}[t]
  \centering
  \includegraphics[width=3in]{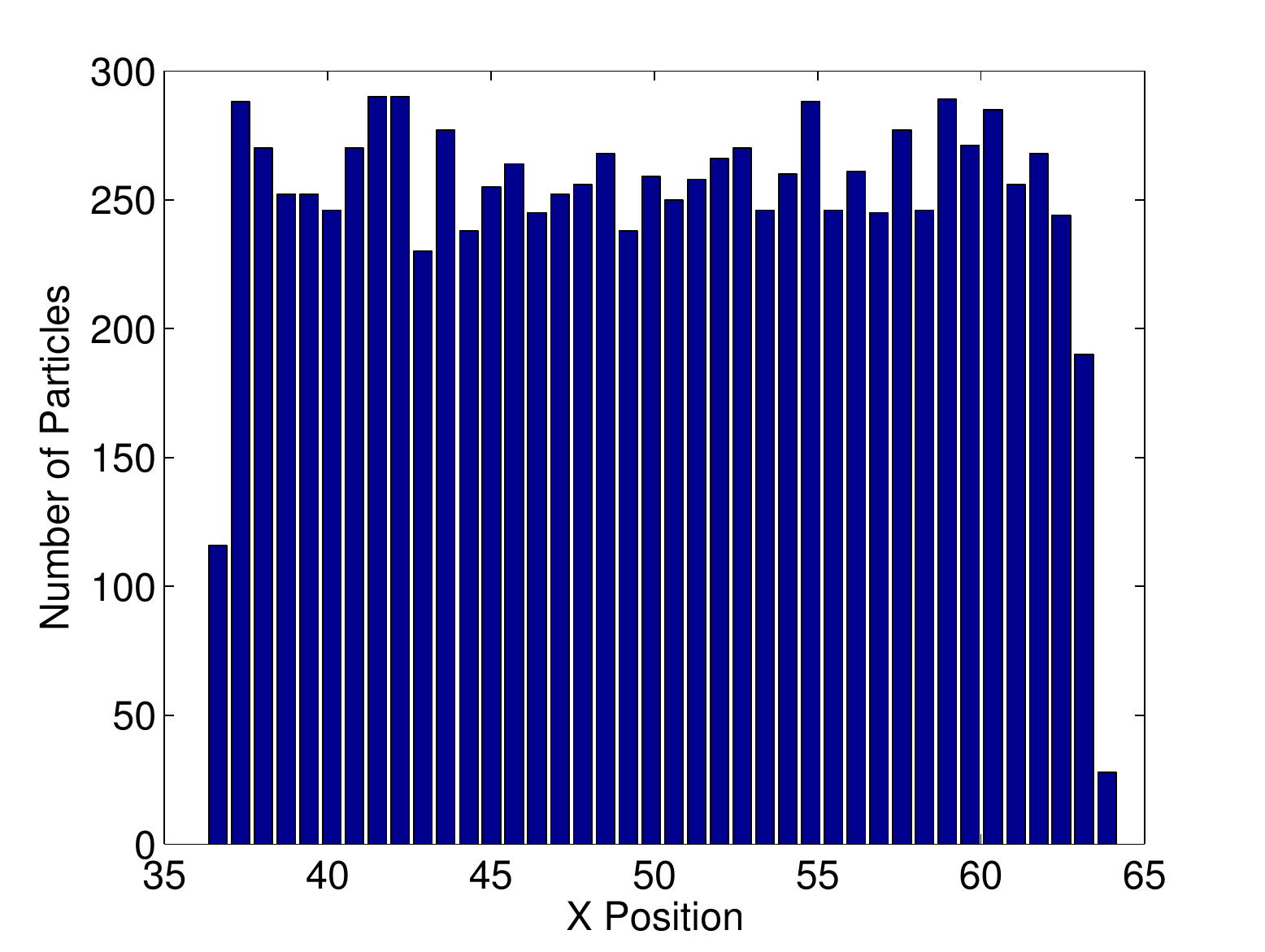}
  \includegraphics[width=3in]{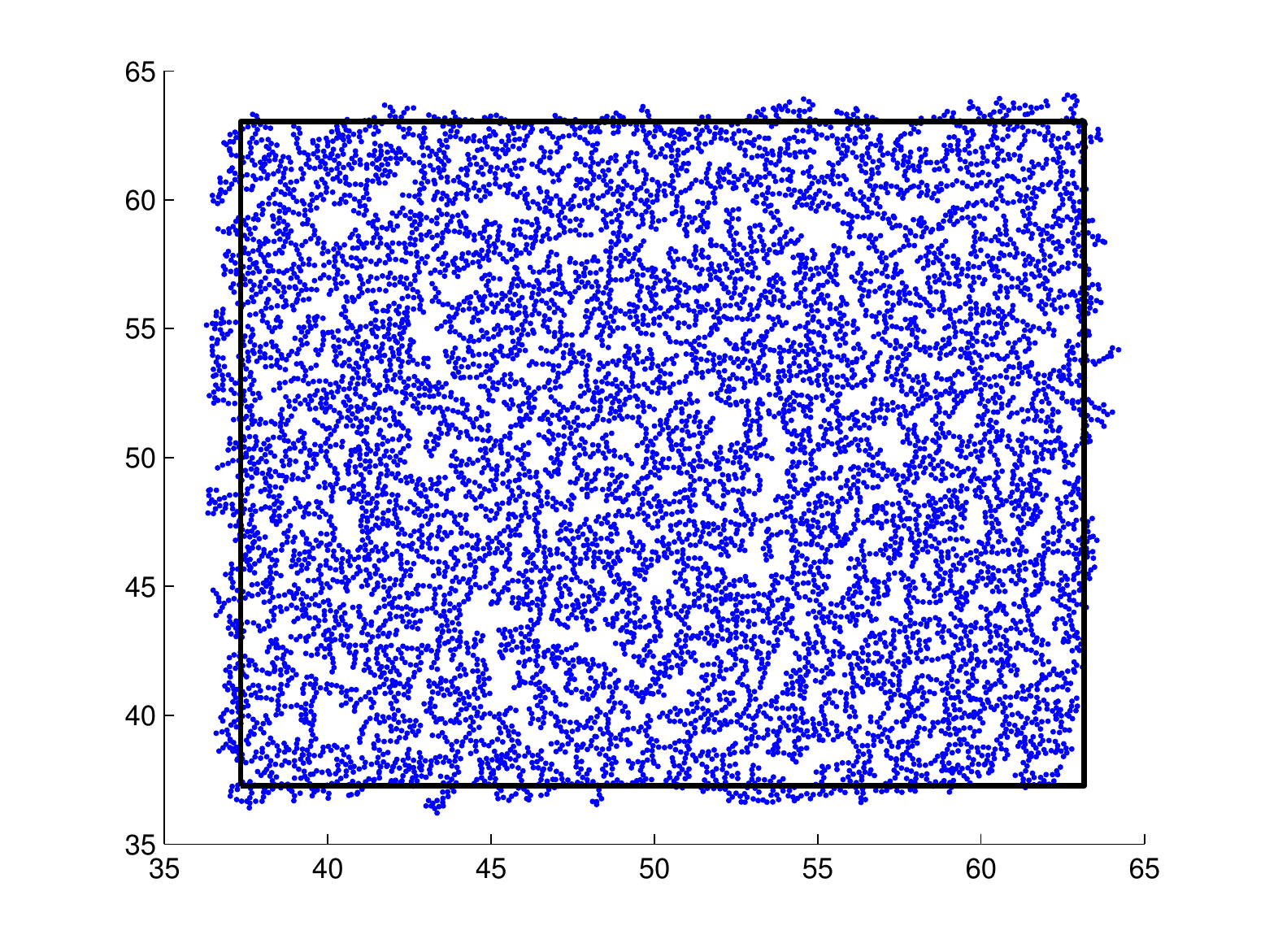}
  \caption{Example 10,000-particle packing (from Algorithm 2) with its associated $x$-position
  histogram and the boundary selected by the method.}
  \label{fig:Histogram}
\end{figure}


The expression in \eqref{eq:PowerSummation} can be computed easily from the solution
of the particle network, so by judiciously choosing $\langle \mytensor{E}\rangle $,
the components of $\mytensor{K}_e$ can be extracted.
In two dimensions, the effective conductivity tensor has three independent components,
so three simulations are sufficient to extract all of the components.
The $K_{11}$ component can be extracted by setting $\langle \mytensor{E}\rangle  = \mytensor{e}_x$.
This corresponds to evaluating the integral for an applied boundary voltage of
$\varphi = -x$.
The remaining tensor components may be similarly extracted by applying specific
potential fields at the boundary and evaluating the summation given in \eqref{eq:PowerSummation}.

\section*{Tests}
The previously described packing algorithms and solution procedures for the current/potential
have been implemented in Matlab.
Algorithm \ref{alg:Algo1} was used to create 50,000 separate 400-particle packings.
In all of these packings, the $F_{11}$ and $F_{22}$ components of the affine transformation
$\mytensor{F}$ equalled $1.0$. 
The $F_{12}$ component that controlled the shearing of the packing ranged between 
0 and 0.5 in increments of 0.01.
Any particles that were sheared out of the original bounding rectangle were reflected to the
other side of the box to return the packing to a rectangular geometry.
Algorithm \ref{alg:Algo2} was used to create 10,000 separate 5,000-particle packings.
In the $\mytensor{B}$ matrix, the $B_{11}$ component remained $1.0$, and
the $B_{22}$ component was varied in $[1.0, 1.9]$ in increments of $0.1$ to influence 
the level of anisotropy of the resulting packings. 
In all simulations, the contact resistance $R_c$ was assigned to be $1$, so it did not
have any affect on the following analyses.
After applying the previously described procedure to each packing to obtain
the effective conductivity tensor and average fabric tensor for each packing,
the data were analyzed to determine how well the results agree with the model's 
prediction for the isotropic magnitude, the deviatoric magnitude, and the direction of conductivity.  
These tests are described next, and thereafter we shall proceed to show how well the lattice model 
performs compared to the existing model, equation \eqref{eq:JH}.

The isotropic behavior of the conductivity can be investigated
by taking the trace of both sides of \eqref{eq:KAtensor2D}.
The average coordination number is the most natural independent
variable when examining the isotropic behavior, so in addition
to taking the trace of both sides of \eqref{eq:KAtensor2D}, both
sides were multiplied by $\det\mytensor{A}$ in order to make the 
right-hand side a single-valued function of $\tr \mytensor{A}$.
This results in \eqref{eq:KAiso}.

\begin{equation} \label{eq:KAiso}
  R_c\,\tr\mytensor{K}\,\det\mytensor{A} = \left(\tr\mytensor{A} - 2\right)\tr\mytensor{A}
\end{equation}

The results of the simulations are plotted together with the analytical
curve given by \eqref{eq:KAiso} in figure \ref{fig:traceBehavior}.
It was found that the analytical solution is usually an upper bound on
the measured conductivity.
This can be explained by the fact that the analytical model was derived
from an idealized system where the chains span a unit cell in a
straight line.
Since the total resistance of a chain is proportional to the number
of contacts in the chain, it follows that the shortest chain between
any two points is the lowest resistance path, and therefore most conductive.
Since the model was derived from a straight-chain idealization, it implies 
an upper bound on the conductivity.  
This logic is less valid in low-coordinated systems, which have many disconnected
groupings of one or two particles;  low-coordinated systems rarely if ever occur 
from Algorithm 2 or in actual carbon black suspension networks.  
In this case, the trace of the system's fabric can be less than 2 but percolating 
chains may still exist to produce small but non-zero conductivity.  
This effect is evident in the figure in the data of Algorithm 1.

\begin{figure}[t]
\begin{center}
 \includegraphics[width=3.2in]{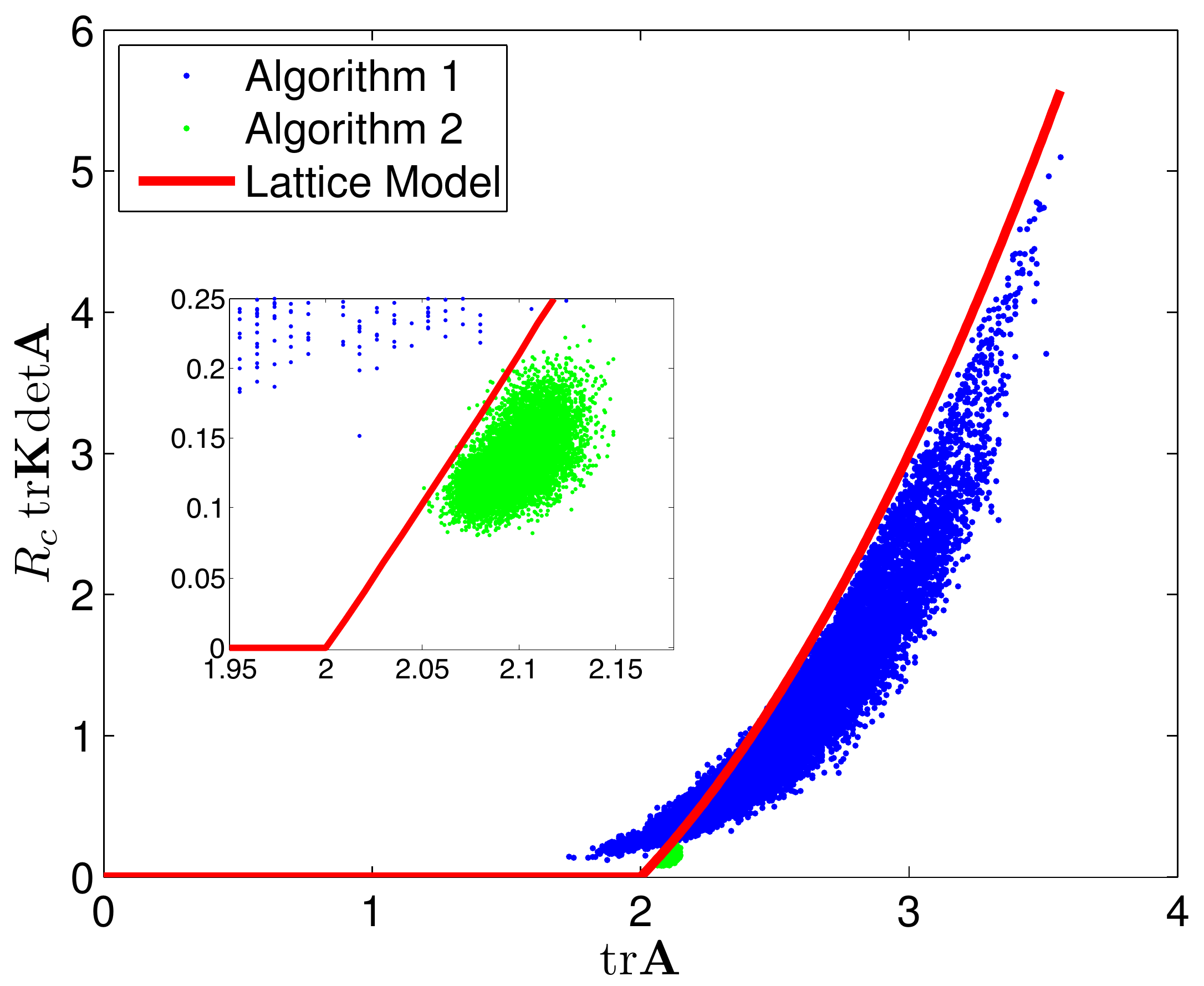}
  \caption{
    Predicted relationship between the (modified) trace of the conductivity and the fabric trace, compared to numerical results of 50,000 packings generated 
    by Algorithm 1 and 10,000 generated under Algorithm 2.  Inset is a zoom-in of the vicinity of $\text{tr}\mytensor{A}=2$.
  }
  \label{fig:traceBehavior}
  \end{center}
\end{figure}

Next, we determine the extent the analytical lattice model predicts
the anisotropy of the conductivity.
To remove the influence of the isotropic behavior, we take the deviator of both
sides of \eqref{eq:KAtensor2D}.
In this case, the most natural independent variable is the magnitude of the
fabric deviator, so the resulting equation was manipulated to be a single-valued
function of this quantity.
After manipulation, \eqref{eq:KAtensor2D} can be written as \eqref{eq:deviatorQty}.
\begin{equation} \label{eq:deviatorQty}
  R_c\,\mytensor{K}_0 : \frac{\mytensor{A}_0}{|\mytensor{A}_0|} 
    \left( \frac{\det\mytensor{A}}{\tr\mytensor{A}-2}\right) = |\mytensor{A}_0|
\end{equation}
where a subscript $0$ denotes the deviator of the tensor, and the term $\frac{\mytensor{A}_0}{|\mytensor{A}_0|}$ is commonly referred to as the direction or sign of the tensor $\mytensor{A}_0$. 
The left hand side of this was plotted against $|\mytensor{A}_0|$ to test
the predictive power of the model.
It can be seen in figure \ref{fig:deviatorQty} that, although there is a 
large amount of noise in the measurements,
the model captures the mean behavior very closely.

\begin{figure}[t]
  \centering
  \includegraphics[width=3in]{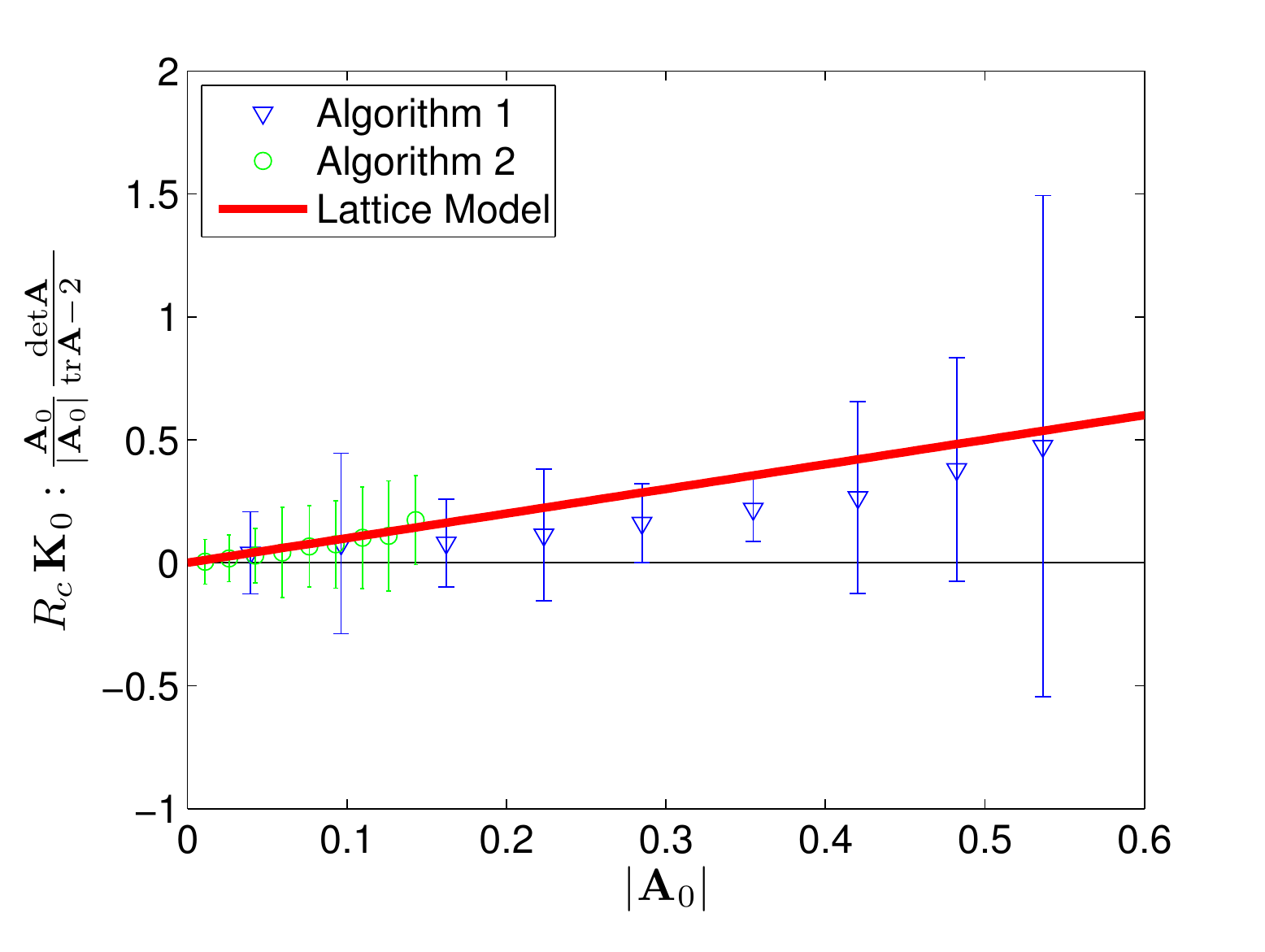}
  \caption{
    Predicted relationship between effective magnitude of the anisotropy 
    of the conductivity and the invariants of the fabric. 
    Error bars show $\pm$ one standard deviation. 
  }
  \label{fig:deviatorQty}
\end{figure}

The final prediction that must be examined is the notion of codirectionality.
The analytical model in \eqref{eq:KAtensor2D} predicts that the fabric
and conductivity tensors have the same eigenvectors.
To examine this, the angle difference between the fabric and conductivity
deviators was calculated, which is equivalent to the (signed) angle between 
the eigenvectors corresponding to the largest eigenvalues of the two tensors, 
denoted $\mathbf{e}_K$ and $\mathbf{e}_A$.
The deviators were chosen because, in 2D, the eigenvector corresponding
to the positive eigenvalue can be unambiguously chosen.
The probability density function of the angle difference as a function of
$\Delta \theta$ is plotted in figure \ref{fig:AngleFig}.
It can be seen that this distribution is symmetrically centered around zero, 
indicating that the fabric and conductivity are strongly codirectional.

\begin{figure}
  \centering

  \includegraphics[width=.34\textwidth]{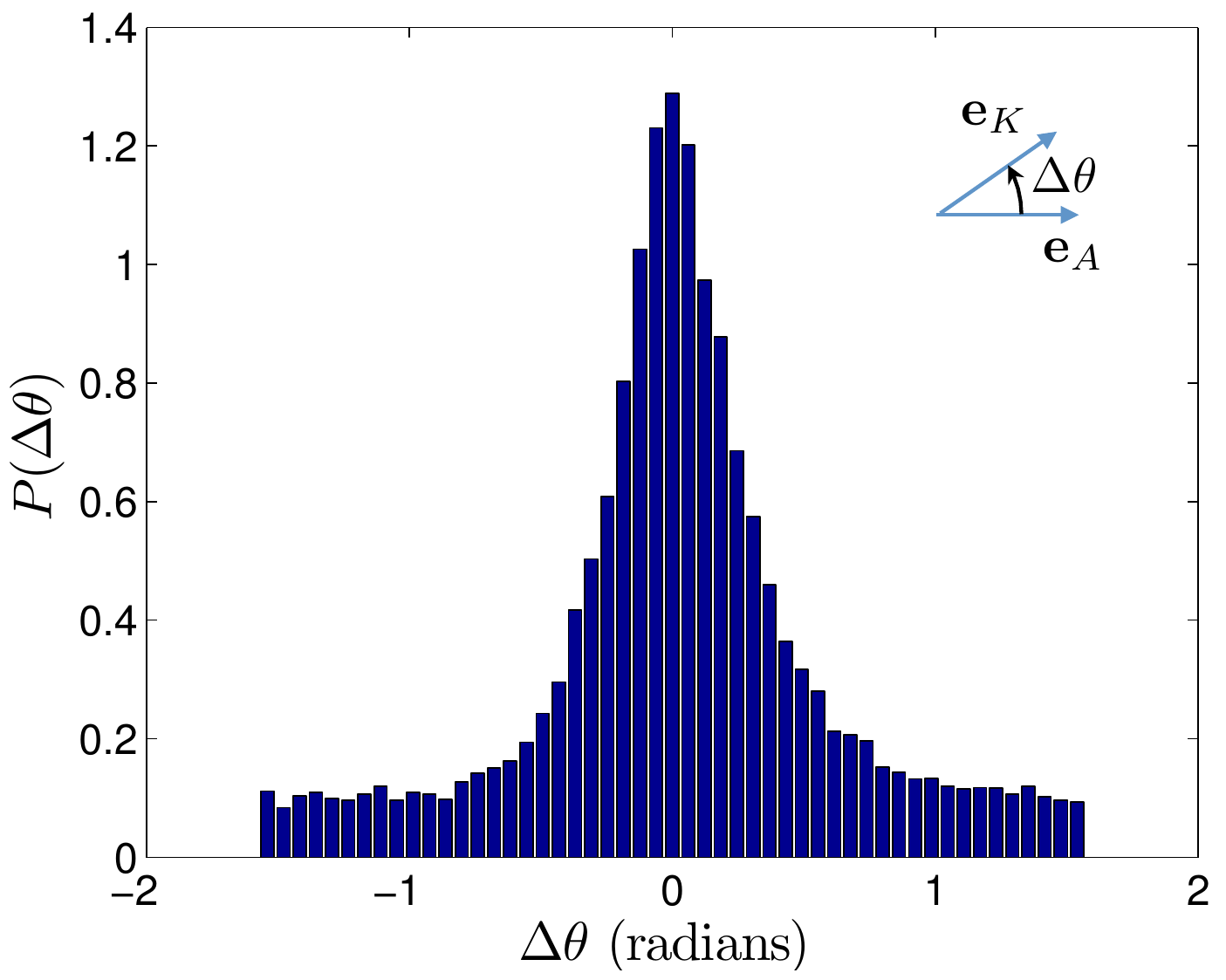}  
    \caption{
    PDF of angle differences are distributed around zero, indicating codirectionality 
    of the fabric and conductivity tensors, as predicted by the analytical model, 
    i.e. \eqref{eq:KAtensor2D}. 
  }
  \label{fig:AngleFig}
\end{figure}

Finally, we also compared the lattice model, \eqref{eq:KAtensor2D}, 
to the existing model by Jagota \& Hui\cite{Jagota1990}
shown in \eqref{eq:JH}.
For the same 60,000 packings generated using both packing algorithms, we computed the relative error of the prediction of the trace and the determinant of the conductivity
using each model and plotted the results in figures \ref{fig:traceError} and \ref{fig:detError}.
In every case, we found that the new lattice model predictions were closer
to the true values from the numerical experiments than the previous
model by Jagota \& Hui.  On the other hand, the Jagota \& Hui model maintains a strong upper bound on both invariants of the conductivity tensor, whereas the lattice model is not strictly an upper bound, as previously discussed. 

\begin{figure}[t]
  \centering
  \includegraphics[width=1.77in,clip=true]{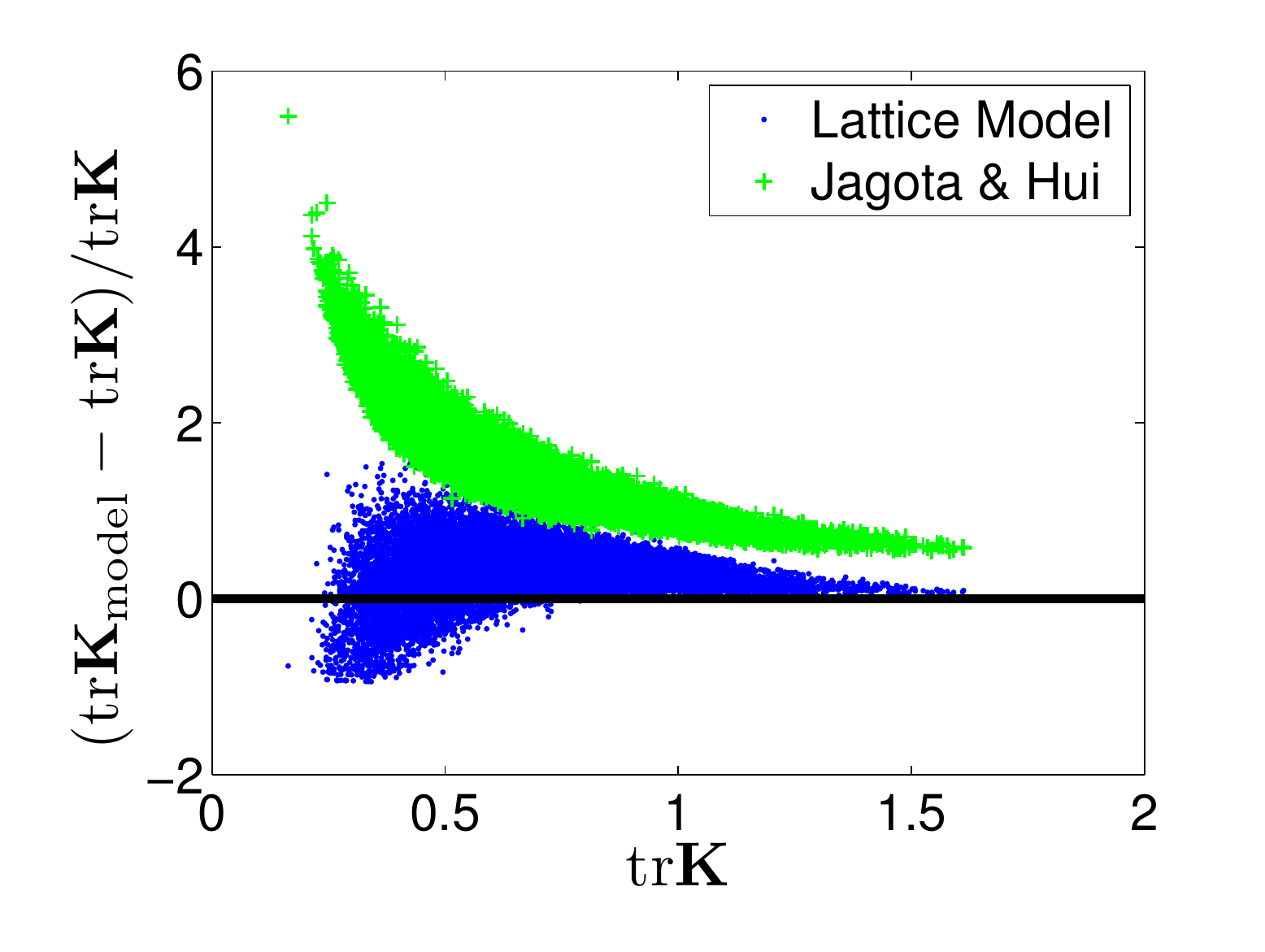}
  \hspace{-0.2in}
  \includegraphics[width=1.77in,clip=true]{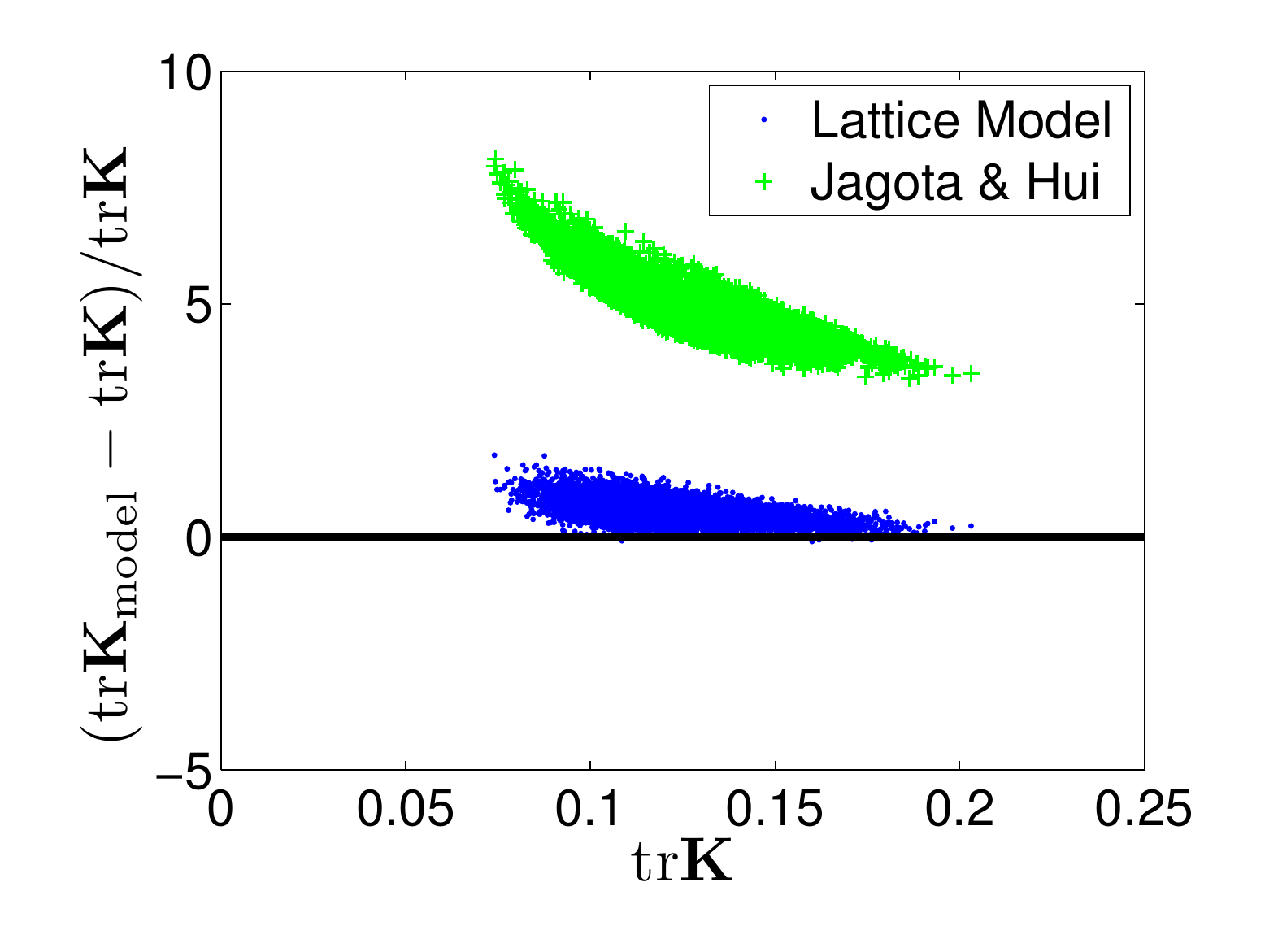}
  \caption{
    Plot of the relative error of the trace of conductivity.
    (Left) Relative error from packings created with Algorithm \ref{alg:Algo1}.
    (Right) Relative error from packings created with Algorithm \ref{alg:Algo2}. 
  }
  \label{fig:traceError}
\end{figure}

\begin{figure}[t]
  \includegraphics[width=1.77in,clip=true]{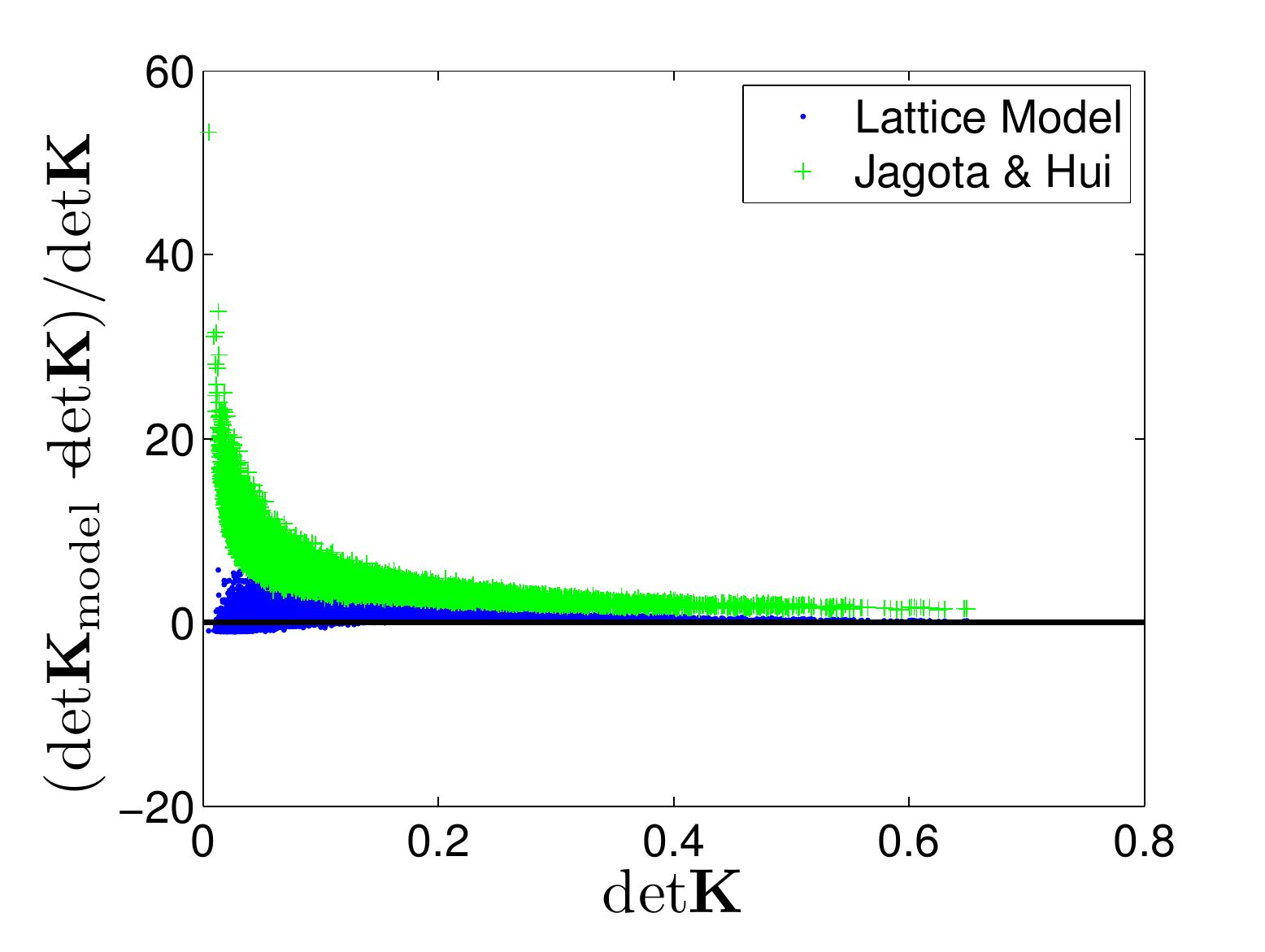}
  \hspace{-0.2in}
  \includegraphics[width=1.77in,clip=true]{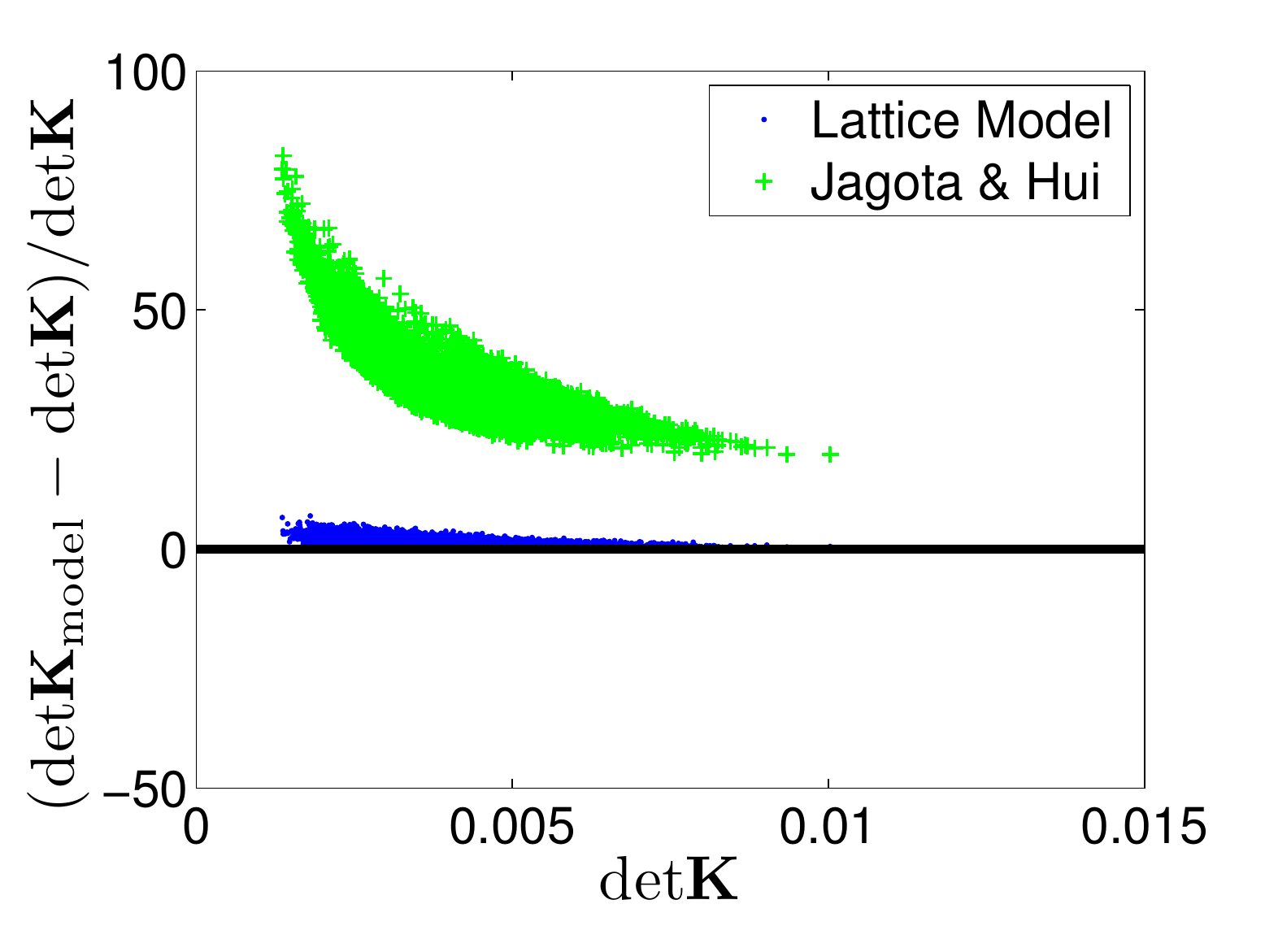}
  \caption{
    Plot of the relative error of the determinant of conductivity.
    (Left) Relative error from packings created with Algorithm \ref{alg:Algo1}.
    (Right) Relative error from packings created with Algorithm \ref{alg:Algo2}.
  }
  \label{fig:detError}
\end{figure}

\section*{Discussion and Conclusions}
In this paper we have derived and tested a new model relating the structure of a
packing of particles to its tensorial electrical conductivity.
The assumptions implicit in the model are that the suspending medium is a
perfect insulator and that electrical resistance arises only at particle contacts.  
The structural measurement used was the fabric tensor, and the model arises from a 
straightforward analysis of a representative problem involving a lattice structure.
The resulting model takes a nonlinear functional form, and was tested multiple ways 
against numerical simulations of many thousands of random particle packings.  The 
agreement in its predictions of the various scalar properties and tensorial orientation 
is significant, especially in light of the simplistic nature of the fabric tensor 
being the sole independent variable for the model.   
In our tests, the lattice model's accuracy was shown to be higher than an existing 
conductivity model, a model which requires more structural input data than the lattice model.  
While it is definitely possible to write a more accurate model by including dependences on more 
structural variables --- some of our data spread is due to the finite 
nature of the datasets, but some is surely due to modeling error --- the current simplicity of the lattice model is an advantage for its usage in engineering applications involving flowing suspension networks.  
Modeling frameworks for the evolution of anisotropy tensors in flowing media have been 
developed over the last decades\cite{Hand1962,Frederick2007,Radjai2012}; keeping our model in terms of fabric, then, suggests a path to the simulation of simultaneous flow and current transfer fields in nontrivial systems by coupling a fabric evolution rule and a rheology with our conductivity model.  
Such a capability would be key in the targeted application of modeling flow battery systems, which
rely on a flowing conductive suspension that closely resembles
the idealized system that we considered.


\section*{Acknowledgements}
The authors acknowledge support from the Joint Center for Energy Storage Research (JCESR), an Energy Innovation Hub funded by the U.S. Department of Energy, Office of Science, Basic Energy Science (BES).
The authors declare that there are no conflicts of interest.

\bibliographystyle{plain}
\bibliography{citations.bib,extraCitations.bib}

\end{document}